\documentclass{emulateapj}

\shorttitle{Unveiling the three-dimensional structure of galaxy clusters}
\shortauthors{Morandi et al.}

\usepackage{hyperref}
\usepackage{graphicx}
\usepackage{amssymb,amsmath,amsfonts,epsf}

\def\aap{A\&A}

\def\apj{ApJ}

\def\apjl{ApJ}
\def\mnras{MNRAS}
\def\araa{ARA\&A}
\def\aj{AJ}

\def\lesssim{\mathrel{\hbox{\rlap{\hbox{\lower4pt\hbox{$\sim$}}}\hbox{$<$}}}}
\def\gesssim{\mathrel{\hbox{\rlap{\hbox{\lower4pt\hbox{$\sim$}}}\hbox{$>$}}}}
\newcommand{\bu}{{\bf u}}
\newcommand{\br}{{\bf r}}
\def\lesssim{\mathrel{\hbox{\rlap{\hbox{\lower4pt\hbox{$\sim$}}}\hbox{$<$}}}}
\def\gesssim{\mathrel{\hbox{\rlap{\hbox{\lower4pt\hbox{$\sim$}}}\hbox{$>$}}}}
\newcommand{\mathbfit}[1]{\textbf{\textit{#1}}}

\begin{document}

\title{Unveiling the three-dimensional structure of galaxy clusters: resolving the discrepancy between X-ray and lensing masses}

\author{Andrea Morandi\altaffilmark{1}, Kristian Pedersen\altaffilmark{1}, Marceau Limousin\altaffilmark{2,1}}
\altaffiltext{1}{Dark Cosmology Centre, Niels Bohr Institute, University of Copenhagen, Juliane Maries Vej 30, DK-2100 Copenhagen, Denmark - Correspndence to: andrea@dark-cosmology.dk}
\altaffiltext{2}{Laboratoire d'Astrophysique de Marseille, Universit\'e de Provence, CNRS, 38 rue Fr\'ed\'eric Joliot-Curie, F-13388 Marseille Cedex 13, France}


\begin{abstract}
We present the first determination of the intrinsic three-dimensional shapes and the physical parameters of both dark matter (DM) and intra-cluster medium (ICM) in a triaxial galaxy cluster. While most previous studies rely on the standard spherical modeling, our approach allows to infer the properties of the non-spherical intra-cluster gas distribution sitting in hydrostatic equilibrium within triaxial DM halos by combining X-ray, weak and strong lensing observations. We present an application of our method to the galaxy cluster MACS\,J1423.8+2404. This source is an example of a well relaxed object with a unimodal mass distribution and we infer shape and physical properties of the ICM and the DM for this source. We found that this is a triaxial galaxy cluster with DM halo axial ratios $1.53\pm 0.15$ and $1.44\pm 0.07$ on the plane of the sky and along the line of sight, respectively. We show that accounting for the three-dimensional geometry allows to solve the long-standing discrepancy between galaxy cluster masses determined from X-ray and gravitational lensing observations. We focus also on the determination of the inner slope of the DM density profile $\alpha$, since the cuspiness of dark-matter density profiles in the central regions is one of the critical tests of the cold dark matter (CDM) paradigm for structure formation: we measure $\alpha=0.94\pm 0.09$ by accounting explicitly for the 3D structure for this cluster, a value which is close to the CDM predictions, while the standard spherical modeling leads to the biased value $\alpha =1.24\pm 0.07$. Our study proves that is not possible to disprove the manifestation of the DM with $\sim 0.25$ per cent of error of failing to reject the null hypothesis. Our findings provide further evidences that support the CDM scenario and open a new window in recovering the intrinsic shapes and desired physical parameters of galaxy clusters in a bias-free way. This has important consequences in using galaxy clusters as cosmological probes.
\end{abstract}


\keywords{galaxies: clusters: individual (MACS\,J1423.8+2404), gravitational lensing, X-rays: galaxies: clusters}



\section{Introduction}\label{intro}

Clusters of galaxies represent the largest virialized structures in the present universe, formed at relatively late times. They are an optimal place to test the predictions of cosmological simulations regarding the mass profile of dark halos. In this respect their X-ray emission can be successfully used to constrain the mass profile in relaxed systems, where the emitting plasma is expected to be in hydrostatic equilibrium \citep[][]{sarazin1988}.

Besides this method, gravitational lensing provides an opportunity to measure the cluster masses without invoking the assumption of hydrostatic equilibrium implicit in the X-ray based mass determinations \citep{Miralda-Escude1995}. Nevertheless, since lensing is sensitive to the integrated mass contrast along the line of sight, it is natural to expect mass overestimates due to fortuitous alignments with mass concentrations which are not physically related to the galaxy cluster or departures of the DM halo from spherical symmetry \citep{Gavazzi2005}.

So far most studies of both dark matter (DM) and intra-cluster medium (ICM) on galaxy clusters have been limited to the standard spherical geometry, i.e. the length scales along the line of sight are taken as the spherical radii measured on the plane of the sky. First, the spatial resolution of Chandra and XMM-Newton X-ray satellites with their high sensitivity and large collecting area has allowed to resolve the core of the clusters, and have detected departures from isothermality and spherical geometry of the ICM and DM: for example evidence for a flattened triaxial dark matter halo around five Abell galaxy clusters has been reported by \citet{Buote1996}. Second, numerical simulations predicts that DM halos show axis ratios typically of the order of $\sim 0.8$ \citep{wang2009}, disproving the spherical geometry assumption. Third, since lensing is sensitive to the integrated mass contrast along the line of sight, departures from the spherical assumption can justify the long-standing discrepancy between galaxy cluster masses determined from X-ray and strong gravitational lensing observations \citep{Gavazzi2005}, the latter being significantly higher than the former: since cluster mass measurements are sensitive to the assumptions about symmetry, this suggests that clusters with prominent strong lensing features are not spherically symmetric and preferentially elongated along the line of sight increasing the magnitude of the lensing. The galaxy cluster A1689 is a well-studied example of a cluster showing this mass discrepancy \citep[][]{andersson2004,lemze2008,riemer2009,Peng2009}.

While galaxy clusters are one of the key cosmological probes, this hinges on our ability to accurately determine both mass and shapes of clusters. An accurate knowledge of the intrinsic cluster shape is required to constrain structure formation models via observations of clusters. Asphericity in the gas density distribution of clusters of galaxies is crucial in modeling X-ray morphologies and in using clusters as cosmological tools \citep[][]{inagaki1995}. 

Previous attempts to infer the elongation of the ICM in galaxy clusters through a comparison between X-ray and Sunyaev Zel'dovich (SZ) observations \citep[see, for example, ][]{defilippis2005} relied on simple $\beta$-model parameterizations of the surface brightness profiles and on the isothermality of the same ICM. While these works suggested departures of the ICM from the standard spherical modeling, their prescriptions are subject to limitations implied in this modeling of the ICM.

In the last few years the theoretical understanding of DM halos has also improved. While spherically averaged density profiles of DM halos are well fitted by a universal DM density profile \citep{navarro1997}, non-spherical effects on the mass function of dark halos have been studied in detail with numerical simulations \citep{lee1999,jenkins2001,jing2002}. In particular the work of \cite{lee2003} derived analytical solutions for gas embedded in triaxial halos on basis of perturbation theory, and they recovered hydrostatic solutions for the gas density and temperature profiles both for the isothermal and the polytropic equation of state. They also showed that the ICM and DM halos are well approximated by a sequence of concentric triaxial distributions with different eccentricity ratio, which has been inferred analytically.

Given that we observe just the two-dimensional projected properties of galaxy clusters on the plane of the sky and any information along the line of sight is lost in the process of projection, it is in general impossible to derive the intrinsic three-dimensional shape of an astronomical object from a single observation. However, one can overcome the previous limitation by combining observations in different wavelengths. In this paper we propose a new method where we compute general three-dimensional hydrostatic equilibrium solutions of the intra-cluster gas under the gravity of triaxial dark matter halos in a non-parametric way and by combining X-ray and lensing data. Indeed without any assumption for models on the gas density and (deprojected) temperature profile, we assume that the triaxial DM density profile is well described by an analytical model (the generalized Navarro, Frenk \& White model, hereafter gNFW). Thanks to the results of numerical simulations, we know, indeed, sufficiently well the DM physics, which is in fact very simple, only depending on the gravity, unlike the physics of the baryons, which is also affected by sources of non-gravitational energy. Moreover we have removed the observational biases in the determination of the deprojected temperature (and consequently of the mass) by adopting the spectral-like temperature estimator (see Sect. \ref{depr}). In this way we have a bias-free estimate of the deprojected temperature and, therefore, of the cluster mass.

In the present paper we aim at describing both the theoretical aspects of the problem and astrophysical applications on a case study, the galaxy cluster MACS\,J1423.8+2404 (hereafter MACS\,J1423). This is a massive cool-core galaxy cluster at $z=0.539$ which does not show any significant sign of disturbance in its morphology and it appears to be very relaxed \citep{ebeling2001,Kartaltepe2008}. The study of \cite{Limousin2009} proved that there is discrepancy between the 2D X-ray mass recovered under the assumption of the standard spherical geometry and the observed surface mass profile from lensing data, the former being systematically lower than the latter. In this perspective, motivated by the need to improve our knowledge of the 3D physical properties of gas and DM in galaxy clusters and, especially, to understand how the 3D geometry will affect the estimated masses, we believe that MACS\,J1423 is a optimal target for our analysis. In particular, we will combine X-ray, strong (SL) and weak (WL) lensing data in order to infer the geometry, physical parameters and distribution of the ICM and DM, which is a fundamental question in bound systems (galaxies, galaxy clusters, dark matter halos) that form in an expanding universe, as well as to determine of the inner slope of the DM, since the cuspiness of dark-matter density profiles in the central regions is one of the critical tests of the cold dark matter (CDM) paradigm for structure formation \citep[][]{navarro1997}.

The paper is organized as follows. In Sect. \ref{datdd1} we present a model for the three-dimensional structure of galaxy clusters and how to constrain the 3D ICM and DM shape. In Sect. \ref{dataan} we summarize the most relevant aspects of the X-ray data reduction procedure, we outline the method applied to determine the X-ray properties for MACS\,J1423, describing our spectral and spatial analysis. In Sect. \ref{snnen} we apply a joint X-ray and lensing analysis in order to infer the physical properties of MACS\,J1423, while in Sect. \ref{conclusion33b} we discuss our findings. We leave to the appendix the discussion of some technical details of our data reduction procedure.

Hereafter we have assumed a flat $\Lambda CDM$ cosmology, with matter density parameter $\Omega_{0m}=0.3$, cosmological constant density parameter $\Omega_\Lambda=0.7$, and Hubble constant $H_{0}=70 \,{\rm km/s/Mpc}$. Unless otherwise stated, we estimated the errors at the 68.3 per cent confidence level.

\section{Three-dimensional structure of galaxy clusters}\label{datdd1}
In order to infer the model parameters of both the ICM and of the underlying DM density profile, we perform a joint analysis for strong+weak lensing and X-ray data. Indeed the lensing and the X-ray emission both depends on the properties of the DM gravitational potential well, the former being a direct probe of the projected mass profile and the latter an indirect proxy of the mass profile through the hydrostatic equilibrium equation applied on the gas and temperature density. In this perspective we outline the methodology in order to infer physical properties in triaxial galaxy clusters. The general idea is straightforward: a) we start with a triaxial DM density model as described in \cite{jing2002}, which is representative of the total underlying mass distribution and depends on a few parameters to be determined, namely the concentration parameter $c$, the scale radius $r_{\rm s}$ and the inner slope of the DM $\alpha$ b) following \cite{lee2003,lee2004}, we recover the gravitational potential and surface mass profile $k$ of a dark halo with such triaxial density profile c) we solve the hydrostatic equilibrium equation for the density of the ICM sitting in the gravitational potential well previously calculated, in order to infer a theoretical three-dimensional temperature profile $T_{\rm gas}$ in a non-parametric way 4) the joint comparison of $T_{\rm gas}$ with the observed temperature and of $k$ with the observed surface mass give us the parameters of the triaxial DM density model, and therefore all the desired physical properties of ICM and DM triaxial ellipsoids (see Fig. 1).

We start by describing in Sect. \ref{datdd2} the adopted triaxial DM density and gravitational potential model, focusing on the relation between elongation of ICM and DM ellipsoids. In Sect. \ref{datdd2y} we present a toy problem in order to outline the effect of triaxiality on the physical observables.

\subsection{Perturbative Expansion of the Triaxial Halo Potential}\label{datdd2}
In the present study, to parametrize the cluster mass distribution, we consider a triaxial generalized NFW model gNFW \citep{jing2002}:
\begin{equation}\label{aa33344}
\rho(R) = \frac{\delta_{c}\rho_{\rm c, z}}{\left(R/R_0\right)^{\alpha}
\left(1 + R/R_0\right)^{3-\alpha}} ,
\end{equation}
where $R_{0}$ is the scale length, $\delta_{c}$ is the dimensionless characteristic density contrast with respect to the critical density $\rho_{\rm c, z}$ of the universe at the present epoch, and $\alpha$ represents the inner slope of the density profile; $\rho_{\rm c, z}\equiv 3H(z)^2/ 8 \pi G$ is the critical density of the universe at redshift $z$, $H_z\equiv E_z\,H_0$, $E_z \!=\left[\Omega_M (1+z)^3 +  (1-\Omega_M-\Omega_{\Lambda})(1+z)^2 + \Omega_{\Lambda}\right]^{1/2}$, and
\begin{equation}\label{aqrt}
\delta_{\rm c} = \frac{200}{3} \frac{ c^3}{ {{\rm
ln}(1+c)-{c/(1+c)}}} \ ,
\end{equation}
where $c \equiv r_{\rm 200}/r_{\rm s}$ is the concentration parameter, $r_{\rm s}$ is the scale radius, $x \equiv r/r_{\rm 200}$.

The radius $R$ can be regarded as the the major axis length of the iso-density surfaces:
\begin{eqnarray}
\label{eq:isodensity}
R^2= a_r^{2}\left(\frac{x^2}{a_r^2} + 
\frac{y^2}{b_r^2} + \frac{z^2}{c_r^2}\right), \qquad (a_r \ge b_r \ge c_r).
\end{eqnarray}

The ellipsoidal shape of a {\it halo} iso-density surface has been evaluated by defining the two eccentricities: 
\begin{equation}\label{aa4343333}
e_{b_r} \equiv \sqrt{1-\left(\frac{b_r}{a_r}\right)^2} ,
\quad
e_{c_r} \equiv \sqrt{1-\left(\frac{c_r}{a_r}\right)^2} , 
\end{equation}
and $a_r \ge b_r \ge c_r$ implies $e_{b_r} \le e_{c_r}$ . The values of $e^{2}_{\sigma}$ ($\sigma = b_r, c_r$) measure the degree of the deviation of the ellipsoidal iso-density surfaces from the spherical ones along the corresponding principal axis direction.

The gravitational potential of a dark halo with the triaxial density profile (eqn. \ref{aa33344}) can be written as a complex implicit integrals \citep{binney1987}. While numerical integration is required in general to obtain the triaxial gravitational potential, small eccentricities ($e_{b_r}^{2} \le e_{c_r}^{2} \ll 1$) expected in cluster scale halos enable to approach this problem with the perturbative expansion and to infer analytical solutions.

There are mainly two different ways to perform the ellipsoidal perturbation, the equal-volume and the equal-length approach. In the former the perturbed ellipsoids have the same volumes as the unperturbed original spheres; in the latter the perturbed ellipsoids and the unperturbed spheres have the same length scales, i.e. the major (minor) axes of the perturbed ellipsoids will be coincident with the radius of the unperturbed sphere. \cite{lee2003} developed the equal-length perturbation theory in order to provide an ellipsoidal correction to the standard spherical modeling of the observed clusters. While in the standard spherical model the length scale in the direction perpendicular to the line of sight is taken as the spherical tangential radii, i.e. measured on the plane of the sky, in their ellipsoidal model this tangential length scale is no longer the same as the line of sight length. Their equal-length perturbative expansion provides a framework which can be observationally more constrained than the equal-volume one because, for example, in SZ observations the cluster length scales in the direction perpendicular to the line of sight are measured \citep[][]{defilippis2005}.

Here we briefly summarize the findings of \cite{lee2003} relevant for this study. Assuming a triaxial gNFW model for the DM (eqn. \ref{aa33344}), they retrieved the following 1st-order approximation for the gravitational potential $\Phi$:
\begin{equation}\begin{split}
\Phi(\bu) \approx & C\;{{ F_{1}(u) +C\;\frac{e_{b_r}^{2}+e_{c_r}^2}{2}F_{2}(u)}} \\
& {{+ C\;\frac{e_{b_r}^{2}\sin^{2}\theta\sin^{2}\phi +  e_{c_r}^{2} \cos^{2}\theta}{2} F_{3}(u)} ,}
\end{split}\label{aa44425}
\end{equation}
where $\bu \equiv \br/R_{0}$, and $C = 4\pi G\delta_{c}\rho_{\rm crit}R_{0}^{2}$, and the three functions, $F_{1}(u),F_{2}(u)$,and $F_{3}(u)$ are defined as
\begin{equation}\label{aa336}
F_{1}(u) \equiv \frac{1}{\alpha-2}\left[ 1 - \frac{1}{u}\int_{0}^{u}\left(\frac{t}{t+1}\right)^{2-\alpha}dt\right] 
\end{equation}

\begin{eqnarray}\label{aa337}
F_{2}(u)\equiv \frac{1}{\alpha-2}\left[-\frac{2}{3}+\frac{1}{u}\int_{0}^{u}\left(\frac{t}{t+1}\right)^{2-\alpha}dt\right]-\nonumber \\
 \frac{1}{\alpha-2}\left[ \frac{1}{u^3}\int_{0}^{u}\frac{t^{4-\alpha}}{(t+1)^{2-\alpha}} dt  \right]
\end{eqnarray}

\begin{equation}\label{aa338}
F_{3}(u) \equiv \frac{1}{u^3}\int_{0}^{u}\frac{t^{4-\alpha}}{(t+1)^{3-\alpha}}dt .
\end{equation}
In the previous equations $CF_{1}(u)$ represents the spherical contribution to $\Phi(\bu)$, i.e., the gravitational potential for the case of the spherical density profile \citep{makino1998,suto1998}. $CF_{2}$ represents another spherical contribution due to the volume changes of the perturbed ellipsoidal density profiles from the spherical ones, while $CF_{3}$ represents the non-spherical deviation of $\Phi(\bu)$ from the spherical potential and introduces a mild dependency of the potential $\phi$ on $\theta$. $F_{3}$ being an order of magnitude smaller than $F_{1}$ and $F_{2}$, it can be neglected in most of the applications.

\cite{lee2003} found that iso-potential surfaces of triaxial DM halos are still best well approximated as ellipsoids and they model the iso-potential surfaces as triaxial ellipsoids with the rescaled major axis length, $\xi$, and the two eccentricities, $\epsilon_{b_r}$ and $\epsilon_{c_r}$: 
\begin{equation}\label{eqn:potpfile}
\tilde\Phi(\xi) = C 
\left[F_{1}(\xi) + \frac{e_{b_r}^{2}+e_{c_r}^2}{2}F_{2}(\xi) \right] ,   
\end{equation}
and that the eccentricities of iso-potential surfaces $\epsilon^{2}_{\sigma}$ are written in terms of their halo eccentricities 
$e^{2}_{\sigma}$ as
\begin{equation}\label{eqn:ecc18}
\frac{\epsilon^{2}_{\sigma}}{e^{2}_{\sigma}} = 
\frac{(\alpha -2)F_{3}(u)}{1 - (\alpha -2)F_{1}(u) 
- u^{2-\alpha}(1+u)^{\alpha-2}}.
\end{equation}
Notice that $\epsilon_{\sigma}=\epsilon_{\sigma}(e_{\sigma},u,\alpha)$ unlike the constant $e_{\sigma}$ for the adopted dark matter halo profile. In the whole range of $u$, $\epsilon_{\sigma}/e_{\sigma}$ is less than unity ($\epsilon_{\sigma}/e_{\sigma}\sim 0.7$ at the center), and decreases mildly as $u$ increases. This means that the intra-cluster gas is more spherical in the outer volumes than in the center, and it is altogether more spherical than the underlying DM halo. This is intuitively understood because the potential represents the overall average of the local density profile, and also because the gas pressure is isotropic unlike the possible anisotropic velocity ellipsoids for collisionless dark matter.

Strictly speaking, equation (\ref{eqn:ecc18}) is valid only for $e^{2}_{\sigma} \ll 1$. This mathematical demand is just partially in agreement with the ellipticity expected for cluster scale halos: the intermediate-major and minor-intermediate axis ratios of halos are typically of the order of $\sim 0.8$ \citep[][]{wang2009}. In this perspective \cite{lee2003} provided the following fitting formula to infer the correct value $\epsilon_{\sigma}^{\rm num}$ of the eccentricity from the previous analytical expression $\epsilon_{\sigma}$ (eqn. \ref{eqn:ecc18}):
\begin{eqnarray}\label{eqn:fit2}
\frac{\epsilon_{b_r}^{\rm num}}{\epsilon_{b_r}} 
&=& 1 + [0.1 + 0.05\log(1+u)]e_{c_r}^{3} + \nonumber \\
& &[0.2 + 0.03\log(1+u)]e_{b_r}^{3}, \label{eqn:fit1}\\ 
\frac{\epsilon_{c_r}^{\rm num}}{\epsilon_{c_r}} 
&=& 1 + [0.1+0.09\log(1+u)]e_{b_r}^{3} + \nonumber \\& &[0.2+0.03\log(1+u)]e_{c_r}^{3}.
\label{eqn:fit244}
\end{eqnarray}
Equations (\ref{eqn:fit2}) and (\ref{eqn:fit244}) are accurate within $10\%$ errors for $e_{\sigma} < 0.6$, while within $20\%$ errors for $e_{\sigma} < 0.8$.

The iso-potential surfaces of the triaxial dark halo coincide also with the iso-density (pressure, temperature) surfaces of the intra-cluster gas. This is simply a direct consequence of the {\it X-ray shape theorem} \citep{buote1994}; the hydrostatic equilibrium equation (\ref{aa2e}) yields
\begin{equation}\label{eqn:ecc}
\nabla P \times \nabla\Phi = \nabla \rho \times \nabla\Phi = 0 .
\end{equation}

\subsection{Constraining ICM and DM shape: a toy problem} \label{datdd2y}
In order to understand how an elongation of the ICM and DM can be constrained by real observations, we present a simple toy problem and we analyze the effect of triaxiality on the physical observables. We consider an ideal ellipsoidal galaxy cluster with symmetry of revolution along the line of sight (i.e. $e_{b_r}=0, e_{c_r}\neq 0$). The related gravitational potential can be obtained from equation (\ref{aa44425}): 
\begin{equation}
\Phi(\bu)  \approx C\left[ F_{1}(u) \pm \frac{e_{c_r}^{2}}{2} 
\{F_{2}(u) + \cos^{2}\theta F_{3}(u)\}\right],
\label{aa335}
\end{equation}
where the positive and the negative signs in front of $e_{c_r}^{2}/2$ correspond to the oblate [$a_r = b_r > c_r$, $ e_{c_r}= \sqrt{1 - (c_r/a_r)^{2}}$)] and the prolate [$a_r < b_r = c_r$, $ e_{c_r}= \sqrt{1 - (a_r/b_r)^{2}}$] cases, respectively. Note that the $z$-direction is always chosen as the symmetric axis in the above expression.

\begin{figure}[ht]
\begin{center}
\epsscale{1.1}
\plotone{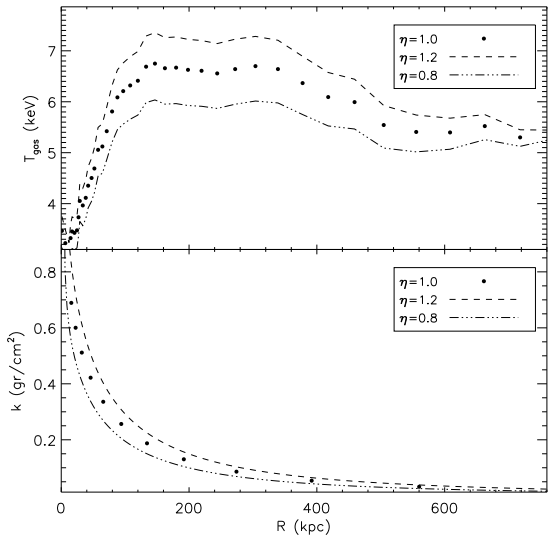}
\caption{The radial profiles for the 3D temperature $T_{\rm gas}(R)$ inferred from the hydrostatic equilibrium equation (upper panel) and the projected mass profiles $k$ (lower panel) for different elongation of the DM halo $\eta_{\rm DM}=\{1,1.2,0.8\}$ (points, dashed and dot-dashed line, respectively). The distance of the points for the spherical case is representative of the true spatial resolution of X-ray and lensing data for MACS\,J1423.}
\label{entps3}
\end{center}
\end{figure}

We consider an analytical model distribution for the gas density in galaxy clusters, $n_e(r)={n_0\; (r/r_c)^{-\varepsilon} {(1+r^2/r_c^2)^{-3/2 \, \beta+\varepsilon/2}}}$, with $n_0,\, r_c,\, \varepsilon,\, \beta$ as free parameters. In order to reproduce realistic galaxy clusters, we fix these parameters to the best fit values of the density profile of MACS\,J1423 under the assumption of spherical geometry ($n_0=0.20\, {\rm cm^{-3}};\; r_c=21.7 \,{\rm kpc}, \; \varepsilon=0.01, \;\beta =0.55$). We assume a spherical NFW model (i.e. $\alpha=1$ in eqn \ref{aa33344}) and fix the concentration parameter and the scale radius to $c=5.55$ and $r_{\rm s}= 253 \,{\rm kpc}$, which are the best expectation values we inferred for MACS\,J1423 in an X-ray-only analysis. We emphasize that we are not trying to reproduce exactly the physical properties of MACS\,J1423, but just a simple toy model for both the ICM and DM, whose parameters can be regarded as realistic for a typical massive galaxy cluster, in order understand how to constrain ICM and DM shape. We considered three different elongations of the DM halo ($\eta_{\rm DM}=\{1,1.2,0.8\}$), corresponding to the case of spherical, prolate and oblate ellipsoid, respectively. From $\eta_{\rm DM}$ we inferred the corresponding elongation of the gravitational potential $\eta_{\rm \phi}=\eta_{\rm \phi}(\eta_{\rm DM},u,\alpha)$, which is ranging from $\{1,1.09,0.90\}$ to $\{1,1.03,0.96\}$ moving from the center towards the X-ray boundary, and we re-calculated the corresponding gas density as $n_{e} \propto \eta_{\rm \phi}^{-0.5}$: indeed what is measured in observations is the emission measure $K\propto \int n^2_{\rm e}\, dV$, so it is pretty straightforward to recover the previous dependency of the gas density on the elongation along the line of sight (see also Appendix~A). In order to infer the deprojected temperature we solved the hydrostatic equilibrium equation (eqn \ref{aa2e}) for the previous values of the elongation of both DM and ICM. Then we calculated the gravitational potential $\phi$ (eqn \ref{aa335}), the gas density $n_{e}$, and the projected mass profile $k$ for the same triaxial DM and ICM halos: the results of our analysis are reported in Fig. \ref{entps3}. Concerning the temperature profiles compared to the spherical case we find an increase (decrease) $\Delta T$ of the temperature for prolate (oblate) ellipsoids, which is small but appreciable in the intermediate region ($\Delta T\sim 0.4-0.6\; {\rm keV \;for} \;R \sim 200-600 $ kpc), but it is smaller near the center and the X-ray boundary. Bearing in mind that the shape of the temperature profile is the same as the underlying gravitational potential thanks to the {\it X-ray shape theorem} (eqn. \ref{eqn:ecc}), the reasonable agreement among the temperature profiles for different values of $\eta_{\rm DM}$ in the inner and outer regions is intuitively understood because: a) the gravitational potential in the outskirts tends to be more spherical b) the temperature has been recovered by solving the hydrostatic equilibrium equation (eqn \ref{aa2e}) via integration of the derivative of the gravitational potential profile $\phi$: it is possibly to show that for a large range of physical radii the relation $T(R,\eta_{\rm DM})/T(R,\eta_{\rm DM}=1) \simeq \eta_{\rm \phi}(R=0)$ roughly holds, therefore it is easy to understand why $\Delta T$ is larger in the central volumes than in the cooling region, where the temperature drops of a factor of 2-3 for a cool-core cluster.

We point out that in Fig. \ref{entps3} the temperature and the projected mass profiles for the spherical case have been represented with points whose distance is representative of the true spatial resolution of X-ray and SL data for MACS\,J1423. Therefore we observe that the spatial resolution is pretty good for both X-ray and lensing data, allowing to accurately measure effects of an elongation along the line of the sight.

Concerning the projected mass profiles we observe that the dependency of the projected mass profile $k$ on the ellipticity of the DM halo is quite straightforward, i.e. $k(R) \propto \eta_{\rm DM}$, so the departures among different projected mass profiles as function of $\eta_{\rm DM}$ are more pronounced in the central regions.

Therefore from this toy model we find that the projected mass profile near the center and the observed temperature profile in the intermediate regions are prone to the value of $\eta_{\rm DM}$: so it easy to understand why both SL and X-ray data can jointly constrain the elongation of ICM and DM ellipsoids along the line of sight, being the former a proxy of the projected mass profiles in the internal regions and the latter of the temperature profile.

As final test, we use this toy model in order to evaluate the systematics involved in our assumption that the triaxial ellipsoid is oriented (oblate or prolate) along the line of sight (see Sect. \ref{depr}). We rotate the previous tree-dimensional ICM and DM ellipsoidal distribution of an angle of 30 degrees respect to the line of sight, we recover the desired two-dimensional physical properties and then we apply our approach. The systematics on the fitting parameters and on the axial ratios have been estimated $\lesssim 3$ per cent, which are within the statistical errors: in particular, the  major-minor axial ratio and the scale radius (the concentration parameter) get slightly underestimated(overestimated). We conclude that this assumption does not strongly affect the estimated parameters and we point out that it is justified in light of the results of \cite{Oguri2009}, who showed that SL clusters with the largest Einstein radii constitute a highly biased population with major axes preferentially aligned with the line of sight increasing the magnitude of lensing.

\section{The dataset and the analysis}\label{dataan}
We focus our analysis on the high-$z$ cool-core galaxy cluster MACS\,J1423.

Motivated by the need to improve our knowledge of the geometry, physical properties and distribution of the ICM and DM in galaxy clusters, we believe that MACS\,J1423 is a optimal candidate for our study, because it is very relaxed, with very good SL, WL and X-ray data.

Here we briefly summarize the most relevant aspects of our data reduction and analysis.

\subsection{Strong and weak lensing analysis}\label{snnen2sl}
We refer to the findings of \cite{Limousin2009}, who presented gravitational lensing and optical properties of MACS\,J1423 based on two bands of HST/ACS observations complemented from ground by K band from the CFHT and Keck spectroscopy. They combined the strong lensing constraints from two spectroscopically confirmed multiply imaged systems with the weak lensing measurements from faint background galaxies via a parametric mass model. Their lensing analysis shows that the cluster has a unimodal mass distribution, suggesting that this an example of relaxed object. From the lensing analysis the cluster looks elongated with a minor-major axial ratio on the plane of the sky of $1.53\pm 0.15$ and position angle of $26\pm 2$ degrees. The 2D projected mass map has been rebinned into elliptical annuli, whose ellipticity, centroid and position angle is the same as that inferred from \cite{Limousin2009}. Then we calculated average values of the elliptical symmetric projected mass profile $k(R)$, $R$ being the minor radius of the 2D elliptical annuli, once we masked out the central 30 kpc, which is affected by the mass distribution of the cD galaxy. We also calculated the covariance matrix $\mathbfit{C}'$ among all the measurements of $k(R)$.

\subsection{X-ray data reduction}\label{laoa}
We have analyzed a Chandra X-ray observation for MACS\,J1423 from the NASA HEASARC archive with a total exposure time of approx. 120 ks. We summarize here the most relevant aspects of the X-ray data reduction procedure for MACS\,J1423. The observation (observation ID 4195) has been carried out using the Back Illuminated S3 chip of ACIS--S. We have reprocessed the event 1 file retrieved from the {\it Chandra} archive with the CIAO software (version 4.1.2) distributed by the {\it Chandra} X-ray Observatory Center. We have run the tool {\tt aciss\_proces\_events} to apply corrections for charge transfer inefficiency, re-computation of the events grade and flag background events associated with collisions on the detector of cosmic rays. We have considered the gain file provided within CALDB (version 4.1.3) in this tool for the data in VFAINT mode. Then we have filtered the data to include the standard events grades 0, 2, 3, 4 and 6 only, and therefore we have filtered for the Good Time Intervals (GTIs) supplied, which are contained in the {\tt flt1.fits} file. We checked for unusual background rates through the {\tt lc\_sigma\_clip}, so we removed those points falling outside $\pm 3\sigma$ from the mean value. Finally, we filtered ACIS event files on energy selecting the range 300-9500 keV and on CCDs, so as to obtain an events 2 file.

\subsection{X-ray spatial and spectral analysis}\label{sp}
We outline the methodology of spatial and spectral analysis in triaxial galaxy clusters. The general idea is to measure the gas density profile in an non-parametric way from the surface brightness recovered by a spatial analysis, and to infer the observed projected temperature profile by a spectral analysis.

The images have been extracted from the events 2 files in the energy range ($0.5-5.0$ keV), corrected by the exposure map to remove the vignetting effects, by masking out the point sources. We determined the centroid ($x_{\rm c},y_{\rm c}$) of the surface brightness by locating the position where the X and Y derivatives go to zero, which is usually a more robust determination than a center of mass or fitting a 2D Gaussian if the wings in one direction are affected by the presence of neighboring substructures. We constructed a set of $n$ ($n= 43$) elliptical annuli of minor radius $r_{m}$ around the centroid of the surface brightness and with ellipticity ${\epsilon_{b'}(r)}$ fixed to that predicted from the ellipticity ${e_{b'}(r)}$ of the DM halo from SL data (see Sect. \ref{depr}, Sect. \ref{datdd2} and eqn. \ref{eqn:ecc18}). We discuss in Sect. \ref{snnen2} to which degree this assumption is consistent with the observed elongation of the X-ray isophotes. The minor radius of each annulus has been selected out to a maximum distance $R_{\rm spat}=890 \,{\rm kpc}$, selecting the minor radii according to the following criteria: the number of net counts of photons from the source in the (0.5-5.0 keV) band is at least 200-1000 per annulus and the signal-to-noise ratio is always larger than 2. The background counts have been estimated from regions of the same exposure, which are free from source emissions.

The spectral analysis has been performed by extracting the source spectra from $n^*$ ($n^*=8$) elliptical annuli of minor radius $r^*_{m}$ around the centroid of the surface brightness and with ellipticity equal to that predicted from the ellipticity ${\epsilon'_{b}(r)}$ of the DM halo from SL data (see above). We have selected the minor radius of each annulus out to a maximum distance $R_{\rm spec}=880 \,{\rm kpc}$, according to the following criteria: the number of net counts of photons from the source in the band used for the spectral analysis is at least 2000 per annulus and corresponds to a fraction of the total counts always larger than 30 per cent.

The background spectra have been extracted from regions of the same exposure, and we have checked for systematic errors due to possible source contamination of the background regions. This is done considering also the ACIS ``blank-sky" background files: we have extracted the blank sky spectra from the same chip regions as the observed cluster spectra and scaled the blank sky spectrum level to the corresponding observational spectrum in the 9-12 keV interval, where very little cluster emission is expected. Then we have applied the aspect solution files of the observation to the background dataset by using {\tt reproject\_events}, so as to estimate the background for our data. We have compared the two methods of background subtraction and we found that the difference between the final results (e.g. the temperature) is negligible. In the following we have used only the results obtained using the local background.

All the point sources have been masked out by both visual inspection and the tool {\tt celldetect}, which provide candidate point sources. Then we have calculated the redistribution matrix files (RMF) and the ancillary response files (ARF) for each annulus: in particular we have used the tools {\tt mkacisrmf} to calculate the RMF, and the tool {\tt mkarf} to derive the ARF of the regions.

For each of the $n^*$ annuli the spectra have been analyzed by using the package XSPEC \citep[][version 11.3.2]{1996ASPC..101...17A} after grouping the photons into bins of 20 counts per energy channel (using the task {\tt grppha} from the FTOOLS software package) and applying the $\chi^2$-statistics. The spectra are fitted with a single-temperature absorbed MEKAL model \citep{1992Kaastra, 1995ApJ...438L.115L} multiplied by a positive absorption edge as described in \cite{vikhlinin2005}. This procedure takes into account a correction to the effective area consisting in a 10 per cent decrement above 2.07 keV. The fit is performed in the energy range 0.6-7 keV (0.8-5 keV for the outermost annulus only) by fixing the redshift to the value obtained from optical spectroscopy and the absorbing equivalent hydrogen column density $N_{\rm H}$ to the value of the Galactic neutral hydrogen absorption derived from radio data \citep{1990ARA&A..28..215D}. We consider three free parameters in the spectral analysis for the $m-$th annulus: the normalization of the thermal spectrum $K_{\rm m} \propto \int n^2_{\rm e}\, dV$, the emission-weighted temperature $T^*_{\rm proj,m}$; the metallicity $Z_{\rm m}$ retrieved by employing the solar abundance ratios from \cite{grevesse1998}.

The global (cooling-core corrected) temperature $T_{\rm ew}$ has been derived in an elliptical region of ellipticity ${\epsilon_{b'}(r)}$ and minor radius $R$, with $50 {\rm \, kpc}<R<R_{\rm spec}$, centered on the symmetrical center of the brightness distribution. We obtained a global cooling-core corrected temperature $T_{\rm ew}=6.34^{+0.24}_{-0.23}$ keV and an abundance of $0.52\pm0.07$ solar value for a reduced $\chi^2$ of 0.88 (356 degrees of freedom).

We classify this cluster as a strong cooling core source (SCC) \citep[][]{morandi2007b}, i.e. the central cooling time $t_{\rm cool}$ is significantly less than the age of the universe $t_{\rm age, z}$ at the cluster redshift ($t_{\rm cool}/t_{\rm age, z} <0.1$): we estimated a $t_{\rm cool}\simeq 6\times 10^8$ yr. As other SCC sources, MACS\,J1423 show very low central temperature ($\sim 2$ keV) and a strong spike of luminosity in the brightness profile. The temperature profile is very regular, suggesting a relaxed dynamical state (see upper panel of Fig. \ref{entps332333}).

\subsection{X-ray spectral deprojection analysis in triaxial ellipsoids}\label{depr}

To measure the pressure and gravitating mass profiles in our clusters, we deproject the projected physical properties obtained with the spectral and spatial analysis by using an updated and extended version of the technique presented in 
\cite{morandi2007a}. Here we summarize briefly the main characteristics of the adopted technique: (i) the electron density $n_e(r)$ is recovered both by deprojecting the surface brightness profile and the spatially resolved spectral analysis obtaining a few tens of radial measurements in elliptical annuli; (ii) once a functional form of the DM density profile $\rho=\rho({\bf {r, q}})$ is assumed (i.e. a gNFW model, see eqn. \ref{aa33344}), where ${\bf q}=(q_1, q_2, ... \, q_h)$ are free parameters of the DM analytical model, and the gas pressure $P_0$ at $R_{\rm spec}$ is kept frozen to the observed value, the deprojected gas temperature $T({\bf q})$ is obtained by integration of the hydrostatic equilibrium equation: 
\begin{equation}\label{aa2e}
\nabla P = -\rho \nabla \phi
\end{equation}
where $\mu=0.6$ is the average molecular weight, $m_p$ is the proton mass, $P=\rho k T/\mu m_{\rm p}$ is the gas pressure, $\rho $ is the gas density and $\phi$ is the gravitational potential of the cluster. So $T({\bf q})=P({\bf q})/n_{\rm gas}$ expressed in keV units.

This approach is very powerful in order to infer the 3D temperature $T({\bf q})$, gas and DM mass profile, because it does not require any ``real" spectral analysis for $T({\bf q})$, which could suffer of the poorness of the statistics and would need at least $\sim$ 2000 net counts per annulus: we can determine the density in elliptical annuli even with very small counts ($\sim 200 - 1000$). In other words we have an improvement (of about one order of magnitude) of the spatial resolution in the spectral analysis: indeed we remember that the total number $n$ of elliptical annuli where we performed the spatial analysis is $n= 43$, while the total number $n^*$ of those where we performed the spectral analysis is $n^*=8$.

Given that in eqn. \ref{eq:isodensity} the condition $a_r \ge b_r \ge c_r$ must always hold and the axial ratio of the elliptical DM halo are free parameters to be determined, i.e. unknown a priori and that might change orientation for each of the MCMC simulations (see below), we redefine new reference axes $(a',b', c')$, which are simply azimuthally and longitudinally rotated of $\{0,\pm 90\}$ degrees respect to the reference axes $(a_r,b_r,c_r)$, and their orientation is frozen along the minor and major axis on the plane of the sky and the axis along the line of sight, respectively. Hereafter, unless otherwise stated, we will always refer implicitly to the reference axes $(a',b', c')$, therefore $e_{b'}$ ($\epsilon_{b'}$) and $e_{c'}$ ($\epsilon_{c'}$) will refer to the eccentricity on the plane of the sky and along the line of sight, respectively\footnote{$e_{b'}=\sqrt{1-(b'/a')^2}$; $e_{c'}= \sqrt{1 - (c'/a')^{2}}$) for oblate ellipsoids, $ e_{c'}= \sqrt{1 - (a'/c')^{2}}$ for prolate ones.}. From Sect. \ref{sp} we remember that the ellipticity $\epsilon_{b'}$ of the elliptical annuli where we performed the spectral and spatial analysis has been fixed to that predicted from the ellipticity $e_{b'}$ of the DM halo from SL data. Therefore, for the triaxial gNFW, we have:
\begin{equation}\label{aa334}
{\bf q}=(c,r_{\rm s},\alpha,e_{c'})
\end{equation}

Concerning an X-ray-only analysis, the comparison of the observed projected temperature profile $T^*_{\rm proj,i}$ (Sect. \ref{sp}) with the deprojected $T({\bf q})$ inferred from hydrostatic equilibrium equation (eqn. \ref{aa2e}), once the latter has been re-projected by correcting for the temperature gradient along the line of sight as suggested in \cite{mazzotta2004}, provides the probability distribution function of model parameters ${\bf q}$ via Markov Chain Monte Carlo (MCMC) algorithm, and therefore of $T({\bf q})$. The results for each of the MCMC simulations will be reported in the reference axes $(a',b', c')$. Even if an X-ray-only analysis might provide information on the desired model parameters, given that we also behave lensing information, we will focus on a joint X-ray+lensing analysis (Sect. \ref{sryen2}).

We point out that, in order to infer the electron density $n_e(r)$, we calculated the electron density $\tilde n_e(r)$ by deprojecting the surface brightness profile in elliptical annuli of eccentricity $\epsilon_{b'}(r)$ and assuming first that an elongation of the ICM ellipsoid halo along the line of sight is the same of the minor axis on the plane of the sky, and then we parametrize our ignorance about it through the following relation:
\begin{equation}\label{aa2hr32e}
n_e(r) = \tilde n_e(r)\; {\eta_{gas,c'}(r)}^{-1/2}
\end{equation}
The proof of the previous eqn. \ref{aa2hr32e} is reported in Appendix~A.

\section{Application of our method on MACS\,J1423}\label{snnen}

\subsection{Joint X-ray+lensing analysis}\label{sryen2}
The lensing and the X-ray emission both depends on the properties of the DM gravitational potential well, the former being a direct probe of the projected mass profile and the latter an indirect proxy of the mass profile through the hydrostatic equilibrium equation applied on the gas temperature and density. In this sense, in order to infer the model parameters, we construct the likelihood performing a joint analysis for WL+SL and X-ray data, to constrain the properties of the model parameters ${\bf q}$ (eqn. \ref{aa334}) of both the ICM and of the underlying DM density profile.

We use the estimate of ${\bf q}$ from an X-ray-only analysis as proposal distribution to start a new MCMC simulation for the joint X-ray and lensing analysis: this makes the calculation faster, because we have a guessed starting value of ${\bf q}$.

\begin{figure}[ht]
\begin{center}
\epsscale{1.1}
\plotone{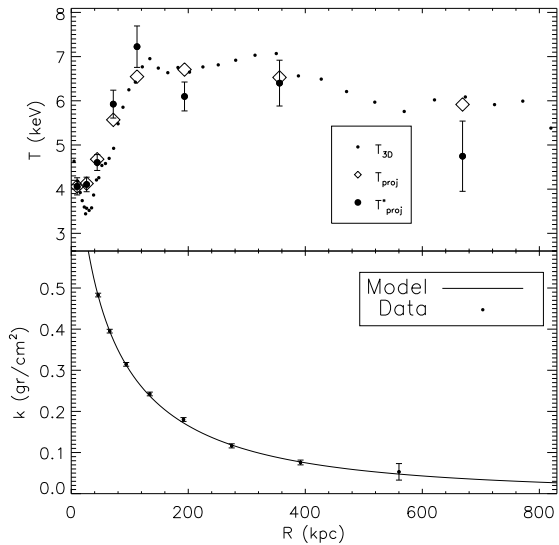}
\caption[]{Example of a joint X-ray and lensing analysis (eqn. \ref{chi2wwf}).
In the upper panel we display the two quantities which enter in the X-ray analysis spectral deprojection analysis (eqn. \ref{chi2wwe}): the observed spectral projected temperature $T^*_{\rm proj,m}$ (big points with errorbars) and the theoretical projected temperature $T_{\rm proj,m}({\bf q})$ (diamonds). We also show the theoretical 3D temperature $T_j({\bf q})$ (points), which generates $T_{\rm proj,m}({\bf q})$ through convenient projection techniques. In the lower panel we display the two quantities which enter in the lensing analysis (eqn. \ref{aa2w2q}): the observed surface mass profile $k_m^*$ (points with errorbars) and the theoretical one $k_m({\bf q})$ (solid line). The distance of the points for both $T_j({\bf q})$ and $k_m^*$ is representative of the true spatial resolution of X-ray and lensing data, respectively.}
\label{entps332333}
\end{center}
\end{figure}

The method works by constructing a joint X-ray+lensing likelihood:
\begin{equation}\label{chi2wwf}
{\mathcal{L}}={\mathcal{L}}_{\rm x}\cdot {\mathcal{L}}_{\rm lens}
\end{equation}
being $\mathcal{L}_{\rm x}$ and ${\mathcal{L}}_{\rm lens}$ the likelihoods coming from the X-ray and SL+WL data, respectively (see below). 

For ${\mathcal{L}}_{\rm x}$ holds the following expression:
\begin{equation}\label{chi2wwv}
{\mathcal{L}}^*_{\rm x}=\frac{\exp \left\{ -\chi^2/2 \right\}} {(2\pi)^{n^*/2} (\sigma_{1}\,\sigma_{2}\,...\,\sigma_{n^*})},
\end{equation}
with $\chi^2$ equal to:
\begin{equation}\label{chi2wwe}
\chi^2= \sum_{i=1}^{n^*} {\frac{{ (T_{\rm proj,i}({\bf q})-T^*_{\rm proj,i})}^2 }
{\sigma^2_{T^*_{\rm proj,i}}  }}\
\end{equation}
being $T^*_{\rm proj,i}$ the observed projected temperature profile in the $i-th$ ring and $T_{\rm proj,i}({\bf q})$ the convenient projection of theoretical 3D temperature $T_j({\bf q})$ in the $j-th$ shell recovered by applying the hydrostatic equilibrium equation (eqn. \ref{aa2e}) and after correcting for the temperature gradient along the line of sight as suggested in \cite{mazzotta2004}. ${\mathcal{L}}_{\rm lens}$ reads:
\begin{equation}\label{aa2w2q}
{\mathcal{L}}_{\rm lens}=\frac{\exp \left\{-\tfrac{1}{2}{[( k_i({\bf q})-k_i^*)]}^{\rm t}\mathbfit{C}^{-1} [( k_i({\bf q})-k_i^*)]\right\}} {(2\pi)^{m^*/2}|\mathbfit{C}|^{1/2}},
\end{equation}
where $\mathbfit{C}$ is the covariance matrix referred to the projected mass profile from lensing data including systematic effects (see below), $|\mathbfit{C}|$ indicates the determinant of $\mathbfit{C}$, $k_i^*$ is the $i-th$ observed measurement of the projected mass profile in the $i-th$ elliptical annulus, $k_i({\bf q})$ the theoretical projected mass profile retrieved by our triaxial DM model (eqn. \ref{aa33344}), $m^*$ the total number of annuli.

For the covariance matrix $\mathbfit{C}$, the below expression holds:
\begin{equation}\label{aartb}
\mathbfit{C}={\mathbfit{C}}'\; + {\mathbfit{C}_{\rm sys}}
\end{equation}
being ${\mathbfit{C}}'$ the covariance matrix among the lensing measurements and ${\mathbfit{C}_{\rm sys}}$ the covariance matrix arising from measurements of systematics, i.e. a bias parameter estimator. We parametrize the systematics involved as ${\mathbfit{C}_{\rm sys}}=\sigma^2_{\rm sys}\; {\mathcal{I}}$, where ${\mathcal{I}}$ is the identity matrix and $\sigma_{\rm sys}$ a bias parameter estimator to be determined. $\sigma_{\rm sys}$ represents an example of latent variables (as opposed to observable variables), i.e. variables that are not directly observed but are rather inferred (through a mathematical model) from other variables that are observed and directly measured. We marginalized over $({\bf q}, \sigma_{\rm sys})$ and therefore we have ${\mathcal{L}}={\mathcal{L}}({\bf q}, \sigma_{\rm sys})$.

So we can determine the physical parameter of the cluster, for example the 3D temperature $T_j=T_j({\bf q})$ and the elongation $\epsilon_{c'}(e_{c'})$ of the ICM(DM) along the line of sight, just by relying on the hydrostatic equilibrium equation and on robust results of the hydrodynamical simulations of the DM profiles. In Fig. \ref{entps332333} we present an example of a joint analysis: notice that in the joint analysis both X-ray and lensing data are very well fitted by our model, with a total $\chi^2=11.3$ (15 degrees of freedom).

\begin{figure}[ht]
\begin{center}
\epsscale{1.1}
\plotone{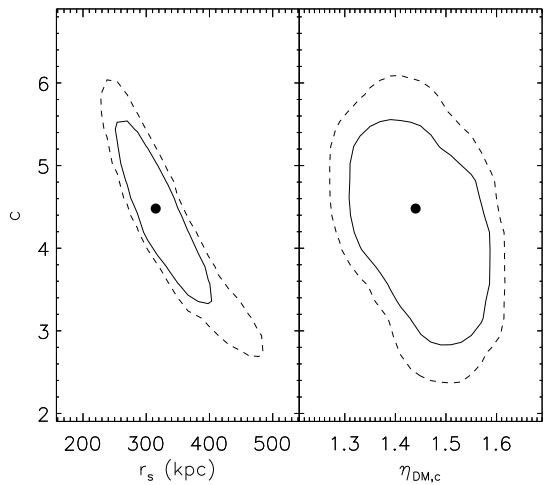}
\caption[]{Marginal probability distribution of $c-r_{\rm s}$ and $c-\eta_{DM,c'}$ (left and  right panel, respectively). The solid(dashed) line represent the 1(2)-$\sigma$ error region, while the big point represents the expectation value.}
\label{entps3xkn}
\end{center}
\end{figure}

In fig \ref{entps3xkn} we present the joint probability distribution of $c-r_{\rm s}$ and $c-\eta_{DM,c'}$, which represent marginal probability distributions of ${\mathcal{L}}({\bf q}, \sigma_{\rm sys})$. We observe that there is an anti-correlation between $c$ and $r_{\rm s}$($\eta_{DM,c'}$). Errors on the individual parameters $({\bf q}, \sigma_{\rm sys})$ have been evaluated by considering average value and absolute mean deviation on the marginal probability distributions of the same parameters.

Finally we computed the total mass enclosed in a ellipsoid ${\mathcal{E}({R_{\Delta},e_{b'},e_{c'}})}$ of major radius $R_{\Delta}$ and of ellipticities $e_{b'},e_{c'}$ as $M({\bf q})({\mathcal{E}({R_{\Delta},e_{b'},e_{c'}})})=\int_{\mathcal{E}} {\rho_{\rm tot}({r},{\bf q}) \, dV}$, where the radius $R_{\Delta}$ corresponds to a given overdensity $\Delta$\footnote{${\mathcal{E}({R_{\Delta},e_{b'},e_{c'}})}$ defines the ellipsoidal volume enclosing an average density $\Delta$ times the critical density of the Universe.}: we considered the case where the overdensity is equal to $2500$.

\begin{figure}[ht]
\begin{center}
\epsscale{1.1}
\plotone{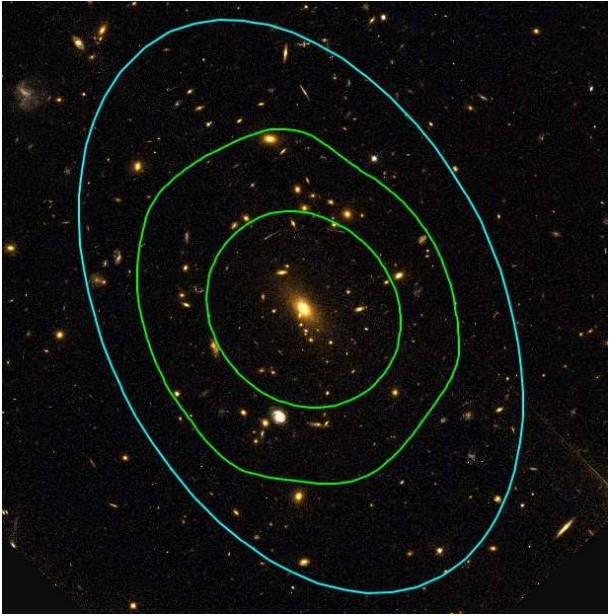}
\caption[]{Optical image of the core of MACS\,J1423 from F555W observation of the NASA/ESA Hubble Space Telescope, with overplotted the projected total mass contours computed from the gravitational lensing analysis (blue line) and from the X-ray surface brightness (green line).}
\label{entps3xc}
\end{center}
\end{figure}

In table \ref{tabdon} we present the best model fit parameters for our analysis of MACS\,J1423. In Fig. \ref{entps3xc} we present an image of the core of MACS\,J1423 from optical (Hubble Space Telescope) observations, with overplotted the projected total mass contours computed from the gravitational lensing analysis (blue line) and from the X-ray surface brightness (green line).

\begin{table*}[htbp]
\begin{center}
\caption{Best model fit parameters of MACS\,J1423. The columns $1-5$ refer to the best fit parameters $c,r_{\rm s},\alpha, \eta_{DM,c'}$ and $\sigma_{\rm sys}$, while the last two columns refer to the mass and radius at $\Delta=2500$, respectively.}
\begin{tabular}{c@{\hspace{.8em}} c@{\hspace{.8em}} c@{\hspace{.8em}} c@{\hspace{.8em}} c@{\hspace{.8em}} c@{\hspace{.8em}}  c@{\hspace{.8em}} }
\hline \\ 
 $c$ & $r_{\rm s}$  & $\alpha$ & $\eta_{DM,c'}$ & $\sigma_{\rm sys}$ & $M_{2500}$ & $R_{2500}$ \\ 
     &     (kpc)    &           &              &${(\rm gr /cm^2)}$ & ($10^{14}M_{\odot}$) & (kpc) \\   
\hline \\ 
  $4.48\pm 0.82$ & $315\pm 65$  &  $0.94\pm 0.09$  &  $1.44\pm 0.07$ & $0.006\pm 0.003$ & $3.41 \pm 0.13$  & $393 \pm 12$ \\   
\hline \\\\
\end{tabular}
\label{tabdon}
\end{center}
\end{table*}

\subsection{An independent test of our method: the elongation of the X-ray isophotes}\label{snnen2}
In order to investigate the validity of our method we observe that the ICM eccentricity $\epsilon_{b'}=\epsilon_{b'}(e_{b'},u,\alpha)$ on the plane of the sky inferred from eqn. (\ref{eqn:ecc18}) can be compared with the observed eccentricity of the X-ray surface brightness contours. We exploit the iterate method proposed by \cite{buote1994} in order to measure the flattening and orientation of the X-ray surface brightness. The parameters obtained from this method, $\eta_{M}$ and $\theta_{M}$, computed within an elliptical region, provide good estimates of the axial ratio $\eta_{M}$ and the position angle $\theta_{M}$ of an intrinsic elliptical distribution of constant shape and orientation. In order to determine these parameters from an image of $P$ pixels having $n_i$ counts in pixel $i$, one computes the moments:
\begin{equation}\label{aa3411}
\mu_{m,n} = {{\frac{1}{N}}  \sum_{i=1}^{p} n_i (x_i-{\stackrel{-}{x}})^m \;(y_i-{\stackrel{-}{y}})^n}\quad  (m,n \le 2), 
\end{equation}
where $N=\sum_{i=1}^{p} n_i$, and $({\stackrel{-}{x}}, {\stackrel{-}{y}})$ is the centroid given by equation $\mu_{1,0}=0$ and $\mu_{0,1}=0$, respectively. Then the axial ratio reads:
\begin{equation}\label{aa3412}
\eta_{M} = {\frac{\lambda_{-}}{\lambda_{+}}}
\end{equation}
and the position angle of the major axis measured north through east in celestial coordinates can be estimated through the following equation:
\begin{equation}\label{aa3413}
\theta_{M} = \arctan{\left({\frac{\mu_{1,1}}{\lambda_{+}^2-\mu_{0,2}}}\right)}+{\frac{\pi}{2}}
\end{equation}
where $\lambda_{\pm} (\lambda_{+}\ge \lambda_{-})$ are the positive roots of the below quadratic equation:
\begin{equation}\label{aa3414}
(\mu_{2,0}-\lambda^2)(\mu_{0,2}-\lambda^2)=\mu_{1,1}^2
\end{equation}

\begin{figure}[ht]
\begin{center}
\epsscale{1.1}
\plotone{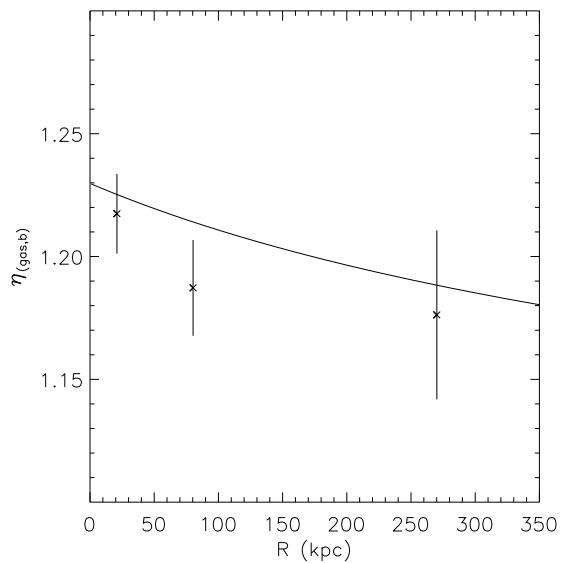}
\caption[]{The radial profiles for the axial ratio $\eta_{\rm{gas},b'}$ on the plane of the sky predicted from eqn. \ref{eqn:ecc18} (solid line) and the observed elongation of the X-ray isophotes (points with errorbars).}
\label{entps332}
\end{center}
\end{figure}

We compute $\eta_{M}$ and $\theta_{M}$ within elliptical annular apertures with increasing inner radii. Given that $\eta_{M}$ is likely a function of the radial distance from the center, the elliptical annuli should correspond more closely to a true isophote since only counts in the immediate vicinity of the isophote are used. We begin by defining a circular aperture for the first annulus ($\eta_{M} = 1$) about the centroid, which has been retrieved on a larger aperture, covering the whole image. Then we compute the appropriate $\mu_{m,n}$ for all the pixels in this elliptical aperture to obtain new values of $\epsilon_{M}$. Defining a new elliptical aperture with these parameters, we iterate until the parameters change by less than appropriate tolerances. These parameters are also used to define the inner radius of the second elliptical annulus, for which we apply the same iterative procedure and so on: indeed we observe that the determination of the desired parameters in each annulus affects the parameters in the more outer annuli. Characterization of the errors in this procedure has been performed via Monte Carlo randomization of uncertainties due to Poisson statistics of the surface brightness. We emphasize that the spread in value of $\eta_{M}$ computed for many different $\eta_{M}$ of the more inner annuli is a measure also of the importance of the systematic errors related to the chosen area of the annuli and departures of the brightness from the elliptical symmetry. Even if we did not find evidence for any position angle twists in the elliptical apertures, during the randomization of uncertainties that might occur because of the strong correlation among the parameters, so we decide to lock the position angle to the value inferred on a larger aperture which covers the whole image and randomize on this value ($\theta_{M}=28.6\pm 1.2$ degrees), which is in good agreement with the value of \cite{Limousin2009} ($\theta_{M}^{\rm SL}=26\pm 2$ degrees).

As pointed out by \cite{buote1994}, for potentials whose shape changes with radius the brightness might not have exactly the same shape as the projected potential, although they have the same three-dimensional shapes. Even if, strictly speaking, just for the case where $\phi$ is stratified on concentric similar ellipsoids the brightness and the projection of $\phi$ have exactly the same shapes, independent of their three-dimensional radial distributions, \cite{buote1994} showed that also for small gradients in ellipticity, the projected shapes should closely approximate the similar ellipsoid case. This kind of systematics was implicitly taken into account in the previous estimate of the errors via Monte Carlo randomization.

In Fig. \ref{entps332} we report a comparison between the theoretical expectation for the axial ratio $\eta_{\rm{gas},b'}$ on the plane of the sky from eqn. \ref{eqn:ecc18} and the observed elongation of the X-ray isophotes. The good agreement confirms the reliability of our method. We remember that our findings just rely on the hydrostatic equilibrium hypothesis and the assumption that the DM follows a triaxial gNFW, therefore the good agreement between observed and predicted axial ratios indicates that the previous underlying assumption holds for this galaxy cluster. In other words, with our method, the desired physical properties are over-constrained by X-ray+lensing observations, providing critical insights into our understanding of clusters, and critical tests of current models (i.e. DM halo model, hydrostatic equilibrium, 3D shapes) for galaxy clusters.

We observe that in Fig. \ref{entps332} a corresponding axial ratio $\eta_{\rm{DM},b'}$ for the DM would be $1.53\pm0.15$, value well above the axial ratio of the gas ($\eta_{\rm{gas},b'}\sim 1.18-1.23$). The probability that the DM axial ratio is in agreement with that of the gas has been discarded with $\sim 0.25$ per cent of error of failing to reject the null hypothesis via the $\chi^2$-test, probing that the DM more elongated than the ICM (see Fig. \ref{entps3xc}). The observed different elongation of halos of the visible matter (i.e. the ICM) constrained by X-ray data and of the gravitational one constrained by lensing observations adds strong evidence, in a very visual way, that the majority of the matter in the system is unseen and under the form of DM for general assumptions regarding the behavior of gravity \citep{clowe2006}.

\section{Discussion}\label{conclusion33b}
The main purposes of this work are to probe the shape of ICM and DM halos and value of the inner DM density profile of MACS\,J1423, two quantities which are intimately connected each other given their degeneracy, as well as to study the discrepancy between X-ray and lensing masses. Here we present the implications of our analysis on the geometry, X-ray and lensing mass measurements (Sect. \ref{conclusion33bbv}) and the CDM scenario (Sect. \ref{conclusion33a}).

\subsection{Resolving the discrepancy between X-ray and lensing masses: probing the 3D geometry of ICM and DM}\label{conclusion33bbv}

Here we outline our findings in probing the 3D shape of ICM and DM, and the systematics involved in using a standard spherical modeling of galaxy clusters. In particular, we will focus on the long-standing discrepancy between X-ray and lensing masses on clusters, showing that this is dispelled if we account explicitly for a triaxial geometry.

MACS\,J1423 is a triaxial galaxy cluster with DM halo axial ratios $\eta_{\rm{DM},b'}=1.53\pm 0.15$ and $\eta_{\rm{DM},c'}=1.44\pm 0.07$, where $\eta_{\rm{DM},b'}$ is the axial ratio of the DM on the plane of the sky inferred from lensing measurements, and $\eta_{\rm{DM},c'}$ the axial ratio of the DM along the line of sight inferred through our analysis (see table \ref{tabdon}). Notice these elongations are statistically significant, i.e. it is possible to disprove the spherical geometry assumption.

Here we are interested in emphasizing the importance of the 3D geometry on the desired physical parameters. If we assume a standard spherical modeling, we obtain $\alpha =1.24\pm 0.07$, $c=2.92\pm 0.77$ and $r_{\rm s}=478\pm 136$, values quit different from those in Table \ref{tabdon}, emphasizing that the elongation of the source along the line of sight does affect the estimates of the physical properties: this proves that the systematics involved in neglecting elongation/flattening of the sources along the line of sight are important. The effects of geometry also may explain the large scatter of concentration parameter and inner slope of the DM found in the literature \citep{Oguri2005}.

We compare our concentration parameter and scale radius with the analysis of \citet{schmidt2007}, who used a circular NFW profile: they found $r_{\rm s}=170^{+30}_{-20}$ kpc and $c=8.27^{+0.72}_{-0.68}$. Again, this larger value of the concentration parameters is significant, showing that departures from sphericity are relevant in retrieving the desired physical parameters. In this perspective large concentration parameter values found in the literature for clusters with prominent strong lensing features can be explained by halos having the major axis oriented toward the line of sight \citep{Corless2009}.

\begin{figure}[ht]
\begin{center}
\epsscale{1.1}
\plotone{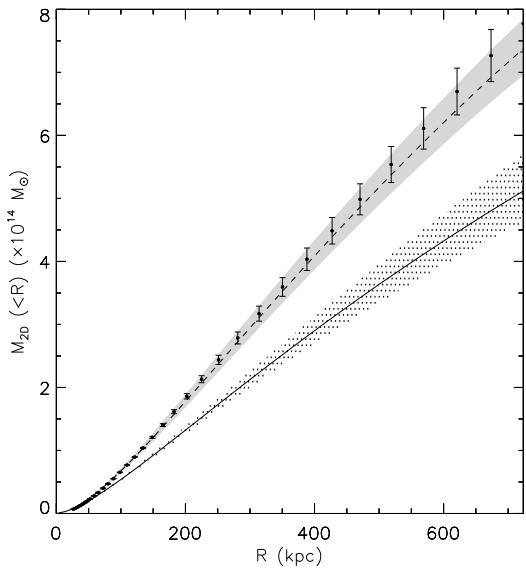}
\caption[]{2D masses enclosed within a circular aperture of radius $R$ from lensing data (points with errorbars), from an X-ray-only analysis under the assumption of spherical geometry (solid line with the 1-$\sigma$ error hatched region), and from a joint X-ray+lensing analysis taking into account the 3D geometry (dashed line with the 1-$\sigma$ error gray shaded region).}
\label{entpsr3}
\end{center}
\end{figure}

Since lensing is sensitive to the integrated mass contrast along the line of sight, departures from the spherical assumption could also justify the disagreement between X-ray and lensing mass profiles found in the literature \citep{Gavazzi2005}. Indeed, in Fig. \ref{entpsr3} we compare the 2D mass enclosed within a circular apertures of radius $R$ for lensing, for an X-ray-only analysis under the assumption of spherical geometry, and from a joint X-ray+lensing analysis taking into account the 3D geometry. We emphasize the good agreement between the masses inferred from lensing and a joint analysis based on triaxial modeling. On the contrary an X-ray-only analysis based on the standard spherical modeling clearly predicts systematically lower masses by $20\div35$ per cent in the radial range $100\div 600$ kpc. Notice that these discrepancies are increasing as we sample larger volumes of clusters: this is understood because of the larger value of $r_{\rm s}$ ($r_{\rm s}=478\pm 136$) for the spherical case respect to the triaxial one ($r_{\rm s}=315\pm 65$). This confirms our insights about the role of the effects of geometry on the physical properties and solve the long-standing discrepancy between X-ray and lensing masses of clusters.

The application of our method on a larger sample of sources will allow to accurately retrieve cluster mass measurements, with important implications in the use of galaxy clusters as cosmological probes through the cluster mass function. As a note of caution, we remember that we focused on a relaxed cool-core object for which the underlying assumptions of our method, i.e. the hydrostatic equilibrium hypothesis and that the DM follows a triaxial gNFW, are likely and has been proved to be tenable (Sect. \ref{snnen2}). While we emphasize the importance of extending our method on other relaxed objects, nevertheless for non-relaxed galaxy clusters our approach might not improve significantly the knowledge of the physical parameters, being our underlying assumptions, in particular that the gas is in hydrostatic equilibrium, not wholly reliable for disturbed galaxy clusters. In this perspective, n-body simulations might address this issue for non-relaxed objects.

\subsection{Probing the inner DM slope}\label{conclusion33a}
While it has long been recognized that N-body studies of large-scale structure formation in cold dark matter (CDM) cosmologies form DM halos whose density profiles are remarkably similar in shape over a wide range of halo mass \citep[e.g.][]{dubinski1991,navarro1997}, a comprehensive physical explanation for the origin of such a profile is still lacking; moreover, when it comes to the shape of the inner density profile of structures, no coherent picture has yet emerged.

\cite{navarro1997} proved a general picture where this universal halo density distribution is characterized by a relatively shallow powerlaw trend in the inner parts, $\rho_{DM}(r) \propto r^{-\alpha}$ , with $\alpha=1$. Subsequently, several studies suggested steeper central cusp \citep[][]{moore1998,moore1999}; with one recent suite of high-resolution simulations \cite{diemand2004} argued for $\alpha =1.16\pm0.14$. \cite{taylor2001} and \cite{dehnen2005} developed a general framework where there is the possibility that $\alpha$ might be shallower than 1 at small radii, while other studies suggested the lack of cusp in the innermost regions \citep{navarro2004}.

On the observational side, efforts have been put on probing the central slope $\alpha$ of the underlying dark matter distribution, through X-ray \citep{ettori2002b,Arabadjis2002,Lewis2003,Zappacosta2006}; lensing \citep{Tyson1998,Smith2001,Dahle2003,Sand2002,Gavazzi2003,Gavazzi2005,Sand2004, Sand2008, Bradac2008,Limousin2008} or dynamics \citep{Kelson2002,Biviano2006}. These studies lead to large scatter in the value of $\alpha$ from one cluster to another, but these determinations rely on the standard spherical modeling of galaxy clusters. Possible elongation/flattening of the sources along the line of sight, as well as the degeneracy of $\alpha$ with other parameters, i.e. $c$ and $r_{\rm s}$, likely affect the estimated values of $\alpha$.

In this perspective, one of the main result of the presented work is to measure a central slope of the DM $\alpha=0.94\pm 0.09$ by accounting explicitly for the 3D structure for MACS\,J1423: this value is close to the CDM predictions from \cite{navarro1997} (i.e. $\alpha=1$), but it is even in better agreement with the more recent numeral simulations of \cite{Merritt2006}, which predicts a slightly shallower inner slope. The value of the concentration parameter is $4.48\pm 0.82$, in agreement with the theoretical expectation from hydrodynamical simulations of \cite{neto2007}, where $c\sim 4$ at the redshift and for the virial mass of MACS\,J1423, and with an intrinsic scatter of $\sim 20$ per cent.

If we carry out a standard spherical modeling, we obtain the biased value $\alpha =1.24\pm 0.07$, value larger than that in table \ref{tabdon}. The different value of $\alpha$ in triaxial and spherical case shows that the systematics involved in neglecting elongation/flattening of the sources along the line of sight are relevant: this likely justifies the large scatter of $\alpha$ found in the literature. 

The lack of a flat core allows us to put a conservative upper limit on the dark matter particle scattering cross section: indeed \cite{yoshida2000} simulated cluster-sized halos and found that relatively small dark matter cross-sections ($\sigma_{\rm dm} = 0.1\, \rm{cm^2\; g^{-1}}$) are ruled out, producing a relatively large ($40 h^{-1}$ kpc) cluster core, which is not observed in our case study.

We observe that the previous DM-only studies neglect the interplay between dark matter and baryons, which are present in observed structures. When it comes to account for the presence of baryons, the exact interplay between dark matter and baryon is not well understood, and different effects, i.e. cooling of the gas and dynamical friction acting on galaxies moving within the DM background, compete in the final determination of the inner slope of the DM. While the former is expected to lead to a more concentrated dark matter density profiles via adiabatic compression \citep{gnedin2004}, the latter can heat up and soften the DM cusp \citep{elzant2004}. In this analysis we did not subtract the mass contribution from the X-ray gas to the total mass. Nevertheless the contribution of the gas to the total matter is small: the measured gas fraction is $0.06-0.07$ in the spatial range $30-400$ kpc, and the slope of the density profile is very similar to that of the DM beyond a characteristic scale $r_c \sim 20-30$ kpc, a self-similar property of the gas common to all the SCC sources \citep[][]{morandi2007b}. This suggest that our assumption to model the total mass as a gNFW is reliable, and therefore a comparison with DM-only studies is tenable. Similar conclusions have been reached by \cite{Bradac2008} and \cite{jesper}.

Both the observed different elongation of halos of the visible and gravitational matter (constrained by X-ray data and lensing observations, respectively) discussed in Sect. \ref{snnen2} and the value of the inner slope of the DM $\alpha$ strongly support the CDM scenario.

\section{Summary and conclusions}\label{conclusion33}
In this paper we have employed a triaxial halo model for the galaxy cluster MACS\,J1423 to extract more reliable information on the three-dimensional shape and physical parameters, by combining X-ray and lensing measurements. We have obtained several significant results.

First, we presented a new method in order to reconstruct the triaxial shape of both DM and ICM. It was demonstrated that the halo triaxiality can cause a significant bias in estimating the desired physical parameters, i.e. concentration parameter $c$, inner slope of the DM $\alpha$ and total mass if a spherical halo model is a priori assumed for the model fitting.

We focused on the implications of our method on the CDM scenario, proving that the value of $c$ and $\alpha$ are in agreement with the CDM predictions \citep{navarro1997}, once we properly accounted for the 3D shape of the cluster. Departures of $c$ and $\alpha$ from the theoretical expectation of the CDM scenario found in the literature can be explained by halos having the major axis preferentially oriented toward the line of sight. In particular, accounting for the 3D geometry allows to resolve the long-standing discrepancy between X-ray and lensing masses in literature.

Then we emphasized the implications of our analysis for DM. Our analysis proves that is not possible to disprove the manifestation of the DM with $\sim 0.25$ per cent of error of failing to reject the null hypothesis.

This confirms our insights about the role of the effects of geometry on the physical properties and allow to solve the long-standing discrepancy between X-ray and lensing masses of clusters. Nowadays the increasing precision of galaxy cluster observations makes the assumption of spherical geometry unlikely to be valid. The triaxiality of dark matter halos and ICM should systematically be taken into account for future analyses on galaxy clusters along the line we suggested. A relevant number of cosmological tests are today based on the knowledge of the mass and shape of galaxy clusters through X-ray measurements by assuming a spherical symmetry. Galaxy clusters play an important role in the determination of cosmological parameters such as the matter density \citep{allen2008}, the amplitude and slope of the density fluctuations power spectrum\cite{voevodkin2004}, the Hubble constant \citep{inagaki1995}, to probe the nature of the dark energy \citep{albrecht2007} and discriminate between different cosmological scenarios of structure formation \citep{gastaldello2007}. It is therefore extremely important to provide a more general modeling in order to properly determine the three-dimensional cluster shape and mass. The application of our method on a larger sample of sources will allow to accurately retrieve cluster mass and shape measurements, with important implications for using clusters as cosmological tools.



\acknowledgments
The Dark Cosmology Centre is funded by the Danish National Research Foundation. KP acknowledges support from Instrument Center for Danish Astrophysics. ML acknowledges the Centre National d'Etude Spatiale (CNES) and the Centre National de la Recherche Scientifique (CNRS) for its support. ML est b\'en\'eficiaire d'une bourse d'acceuil de la Ville de Marseille. We thank Jean-Paul Kneib and Jens Hjorth for useful discussions.




{\it Facilities:} \facility{HST}, \facility{CXO} 




\newcommand{\noopsort}[1]{}

\appendix


\section{Appendix material: Deprojecting spectra and surface brightness in triaxial ellipsoids}\label{apecdepte}
The deprojection technique decomposes the observed X-ray emission of the $i$-th annulus into the contributions from the volume fraction of the $j$-th spherical shells with $j\le i$, by fixing the spectrum normalization (i.e. the emission measure $K_i$) of the outermost shell to the corresponding observed values. We can construct an upper triangular matrix ${\mathcal{V}}_i^j$, where the column vectors ${\mathcal{V}}^1$, ${\mathcal{V}}^2$, ...${\mathcal{V}}^n$ represent the ``effective" volumes, i.e. the volume of the $j$-th shell contained inside the $i$-th annulus (with $j \ge i$) and corrected by the gradient of $n_{e,j}^2$ inside the $j$-th shell \citep[see][for more details on the definition of effective volume instead of the more common geometrical one]{morandi2007a}, so as:
\begin{equation}\label{kk}
K_i \propto \int_{j \ge i} n^2_{\rm e,j}\ dV = 
{\left({\mathcal{V}} \# {\stackrel{\rightarrow}{n_e^2}}\right)}_i\ .
\end{equation}
In the previous equation $\stackrel{\rightarrow}{n_e}\equiv (n_{e,1},n_{e,2},...,n_{e,n})$, being $n^*$ the total number of annuli, having internal (external) radius $r_{\rm in,1}\, ,\, r_{\rm in,2}\, , \, ...\, ,  r_{\rm in,n}$ ($r_{\rm out,1}\, ,\, r_{\rm out,2}\, ,\, ..., r_{\rm out,n} $); $K_i$ is the XSPEC normalization of the spectrum in the $i$-th annulus; the operator $\#$ indicates the matrix product (rows by columns). Notice that the integral $\int_{j \ge i} n^2_{\rm e,j} dV$ is of the order of the emission measure inside the $i$-th ring.\footnote{Hereafter we assume that the index $j$ ($i$) indicates the shell (ring) of the source of radius $(r_{\rm in},r_{\rm out})$. } The inversion of this matrix allows us to determine $n_{e,i}$.

In the standard spherical modeling of the observed clusters the previous shells (annuli) and the matrix ${\mathcal{V}}_i^j$ refer to spherical annuli, while we are interested in extending the (de)-projection schema to concentric triaxial ellipsoids of reference axes $a',b', c'$ with the axial ratios $\eta_{\rm{gas},b'}=b'/a'$ and $\eta_{\rm{gas},c'}=c'/a'$ on the plane of the sky and along the line of sight, respectively. Instead of working on a orthogonal cartesian reference system $(x,y,z)$ we chose another one, ${ (\tilde x,\tilde y,\tilde z)}$, such as $\tilde x=x$, $\tilde y=y/\eta_{\rm{gas},b'}(r)$ and $\tilde z=z/\eta_{\rm{gas},c'}(r)$: in this new reference system it is intuitive to prove that the previous concentric triaxial ellipsoids can be regarded a spherical shells of radius $a'(r)$.

Given the previous axis transformation, the relation between the effective volume matrices $ {\mathcal{V}}$ and $\tilde {\mathcal{V}}$ in the reference system $( x,\ y,z)$ and $(\tilde x,\tilde y,\tilde z)$, respectively, reads:
\begin{equation}\label{kk2}
{{\mathcal{V}}}={{{\rm \bf diag}({\bf \eta_{\rm{gas},b'}})\#{\tilde{\mathcal{V}} \#{\rm \bf diag}({\bf \eta_{\rm{gas},c'}})}}}
\end{equation}
being ${\rm \bf diag}({\bf \eta_{\rm{gas},c'}})$ a diagonal matrix with axial ratios $(\eta_{\rm{gas},c',1},\eta_{\rm{gas},c',2},...,\eta_{\rm{gas},c',n})$ on the main diagonal. Similar considerations hold for ${\rm \bf diag}({\bf \eta_{\rm{gas},b'}})$

Finally, in order to infer the gas density ${\stackrel{\rightarrow}{n_e^2}}$ in concentric triaxial ellipsoids of minor axis $(a'_1,a'_2,...,a'_n)$, given the emission measure ${\bf K}$ in concentric elliptical annuli of minor axis $(a'_1,a'_2,...,a'_n)$, we have:
\begin{equation}\label{kk233}
{\bf K}={\left({{\rm \bf diag}({\bf \eta_{\rm{gas},b'}})\#{\tilde{\mathcal{V}} \#{\rm \bf diag}({\bf \eta_{\rm{gas},c'}})}} \right)}\# {\stackrel{\rightarrow}{n_e^2}}
\end{equation}
Notice that the matrix ${\tilde {\mathcal{V}}}$ refers to spherical annuli of radii $(a'_1,a'_2,...,a'_n)$. From eqn. \ref{kk233} it is easy to proof the dependency of the density on the elongation along the line of sight given by eqn. \ref{aa2hr32e}.

Similar considerations hold for the temperature (de)-projection.

\end{document}